\documentclass[seceq,supplement]{ptptex}

\usepackage{amsmath}
\usepackage{graphicx}
\def\dd{{\rm d}}
\def\la{\langle}
\def\ra{\rangle}
\def\eqd{\overset{\dd}{=}}
\def\hG{\widehat{G}}
\def\E{\mathbb{E}}
\def\Imp{{\rm Im} \,}
\def\Tr{{\rm Tr} \,}
\def\N{{\cal N}}
\title{%        %You can use \\ for explicit line-break.
Anderson model on Bethe lattices: density of states, localization
properties and isolated eigenvalue}

\author{%       %Use \scshape for the family name.
Giulio \textsc{Biroli}$^{1}$, Guilhem \textsc{Semerjian}$^2$ and Marco
\textsc{Tarzia$^3$}
}

\inst{%     %Affiliation, neglected when [addenda] or [errata].
$^1$ IPhT, CEA, 91191 Gif sur Yvette, France \&
CNRS URA 2306\\
$^2$ LPTENS, CNRS UMR 8549,
associ\'ee \`a l'UPMC, 24 Rue Lhomond, 75005 Paris,
France.\\
$^3$ LPTMC-UPMC, UMR CNRS 7600,
4 place Jussieu, 75252 Paris Cedex 05, France.
}

\abst{%         %This abstract is neglected when [addenda] or [errata].
We revisit the Anderson localization problem on Bethe lattices, putting in
contact various aspects which have been previously only discussed separately.
For the case of connectivity
$3$ we compute by the cavity method the density of states and the
evolution of the mobility
edge with disorder. Furthermore, we show that below a certain critical
value of the disorder
the smallest eigenvalue remains delocalized and
separated by all the others (localized) ones by a gap.
We also study the evolution of the mobility edge at the center of the band
with the connectivity, and discuss the large connectivity
limit.}

\begin{document}

\maketitle

\section{Introduction}
The effect of quenched disorder on non-interacting electrons can be
dramatic. The wave-function
completely changes form going from an extended plane wave at zero disorder
to a localized wave
at large disorder. Actually, even an infinitesimal disorder is enough to
induce localization of all eigenstates
in one and two dimensions \cite{50years}. Localized electrons do not lead
to electronic transport except if the system
is coupled to a thermal bath, as a consequence conduction properties
changes completely
in presence of localization.

This phenomenon, called Anderson localization, was discovered in 1958 by
P.W. Anderson \cite{Anderson}.
Since then a huge amount of work has been devoted to it, see for instance
the brief review "Fifty years of Anderson Localization"
\cite{50years} and the
monographs~\cite{book_Lifshits,Luck,book_Pastur,book_Carmona}. Exactly
solvable models naturally played an important role in its understanding.
One dimensional
systems have been studied thoroughly \cite{Luck}. However, in these
systems all states are localized and hence some aspects
of Anderson localization, such as the transition between localized and
extended states, and the critical properties at the mobility edge could
not be studied. All these properties can instead be analyzed on Bethe
lattices (see the next section for a precise definition), which therefore
provide a very
useful benchmark since they are still simple enough to be studied without
making any approximation.

The first analysis of Anderson localization on Bethe lattices was
performed by Abou Chacra, Anderson and Thouless
\cite{Abou1,Abou2} and many other studies followed both in the physics
(see for instance the works~\cite{MiFy,Derrida,Monthus,Alvermann} and references
therein) and mathematics~\cite{Kunz,Klein1,Klein2,ASW} communities.
Our main goal in this work is to combine various point of views on the
problem, and to underline a particular feature of Bethe lattices. In fact, besides the properties of the density of states, the
evolution of the mobility edge with disorder and with the connectivity
$k+1$, we shall discuss the presence of isolated eigenvalues.
Without disorder, the smallest eigenvalue for the tight-binding
Hamiltonian (or the discrete Laplacian) on a Bethe lattice is isolated
from all the others by a gap~\cite{Friedman}.
This eigenvalue plays a very important role in determining the physical
properties of some statistical mechanics models displaying a condensation
transition at low temperature, such as the spherical model \cite{almost}
or non-interacting bosons on a Bethe lattice. The evolution of the
isolated eigenvalue in presence
of disorder is interesting in particular in connection with studies of
disordered bosonic models since it may lead
to somewhat different behaviors than the ones obtained for finite
dimensional systems, where no isolated eigenvalue is found (even though
the relevance of the spectrum of the one-particle kinetic energy is less
obvious in the presence of interactions).
\section{Model and iterative equations}
\label{sec_model}
The Anderson model corresponds to the Hamiltonian:
\begin{equation}
{\mathcal H}=-\sum_{\langle i,j \rangle}t (c_i^\dagger c_j+c_j^\dagger
c_i)+\sum_{i=1}^{N}\epsilon_i c_i^\dagger c_i\,,
\end{equation}
where the first sum runs over the nearest neighbors couples of the
lattice, the second sum runs over all $N$ sites and
$c_i^\dagger, c_i$ are fermionic creation and annihilation operators. For simplicity we consider spinless fermions.
The on-site energies $\epsilon_i$ are independent and identically distributed random
variables, that we shall take in the following uniformly distributed in
the interval $[-W/2,W/2]$.
As the fermions have no interactions, the study of this model amounts to
determine the spectrum of the $N\times N$ random matrix $H_{ij}$ such that
$H_{ij}=-t$ if $i\neq j$ are neighbors on the lattice, $H_{ii}=\epsilon_i$
on the diagonal. In the following we will take $t=1$,
which means that $\epsilon_i$ are measured in units of $t$.

As anticipated in the introduction, the lattice we will focus on is a
Bethe lattice. It is possible, and this is the point of view of almost all
mathematical works, to define it as an infinite regular tree, i.e. a graph
without loop where every vertex has the same degree (that we shall denote
$k+1$ in the following). It is however necessary for the study of
thermodynamical properties of statistical mechanics model (e.g. to define
a free-energy) to consider finite-size versions of the model, with a given
number $N$ of vertices, before taking the thermodynamic limit $N \to
\infty$. There are at least two possible ways to define a finite-size
Bethe lattice. The first one is to consider a finite tree of depth $n$,
that is a tree in which the root vertex (at generation $0$) has $k+1$
offsprings, each of the vertices in the generation from $1$ to $n-1$ has
$k$ offsprings, and finally the leaves of the $n$-th generation have no
descendent. The second way is to consider a random $k+1$-regular graph,
i.e. a graph chosen uniformly at random among all the graphs on $N$
vertices where each of the vertex has degree $k+1$. The properties of such
random graphs have been extensively studied (see ref.~\citen{Wormald} for a
review). It is known in particular that any finite portion of such a graph
is a tree with a probability going to one as $N\to \infty$. The advantage
of the second construction is the absence of any boundary effect: all
sites, for any finite $N$, have $k+1$ neighbors, whereas in the trees of
the first construction the sites on the boundary, i.e. on the leaves of
the $n$-th generation, are asymptotically as numerous as the bulk (volume)
sites and have connectivity one, a pathological situation compared to
finite dimensions. Random regular graphs can be thought of as regular
trees wrapped onto themselves. This absence of boundary is particularly
important for frustrated models, as discussed notably in
ref.~\citen{MePa_Bethe}.

The spectral properties of $H$ are easily studied in terms of the
resolvent matrix,
which is defined as:
\begin{equation}
G_{ij}(z)=\left(\frac{1}{H-z{\mathcal I}}  \right)_{ij}\,,
\end{equation}
where ${\mathcal I}$ is the identity matrix. For example, the density of
eigenvalues $\lambda_\alpha$,
$\rho(E)=\sum_\alpha\delta(E-\lambda_\alpha)/N$, can be obtained as:
\begin{equation}
\rho(E)=\lim_{\eta\rightarrow 0}\frac{1}{N\pi} \Imp \Tr G(E +i \eta)\,.
\end{equation}
Moreover the localization properties of the eigenvectors of $H$ can be
deduced from the behavior of $\underset{\eta\rightarrow 0}{\lim} \, \eta
|G_{ii}(E+i\eta)|^2$ (see for instance refs.~\citen{Thouless,EcoCo}).

When the non-zero off-diagonal elements of $H$ have the structure of a
finite tree it is easy to obtain the diagonal elements of the resolvent
from
\begin{equation}
G_{ii}(z) = \frac{1}{H_{ii}-z-\underset{j\in\partial i}{\sum} H_{ij}^2
G_{j\to i}(z)} \ .
\end{equation}
Here $\partial i$ denotes the set of neighbors of $i$ and $G_{j\to i}(z)$
is the $G_{jj}(z)$ resolvent of a modified matrix in which the edge
between $i$ and $j$ has been removed. These new variables verify the
following recursive equations:
\begin{equation}
G_{i \to j}(z) = \frac{1}{H_{ii}-z-\underset{j'\in\partial i \setminus
j}{\sum} H_{ij'}^2 G_{j'\to i}(z)} \ .
\label{eq_recurs}
\end{equation}
These equations were first obtained in ref.~\citen{Abou1} using
perturbation theory. Another simple way to obtain these equations is to
consider a Gaussian model with a kernel $H-z{\mathcal I}$ (we shall come
back on this in Sec.~\ref{sec_isolato}). More formally one can use
resolvent identities to derive them; this last method has the advantage of
being also valid directly on the infinite Bethe lattice, the matrix $H$
being turned in an operator acting on square-normalizable functions on the
sites of the infinite tree. Note that it has been proven in
ref.~\citen{cvg} that the random regular graph case can be studied using
these recursion equations in the thermodynamic limit, thanks to the local
convergence of random regular graphs to trees.

Let us denote, with a slight abuse of notation, $G(z)$ the diagonal
element of the resolvent for an arbitrary site in the Bethe lattice (the
infinite tree or the thermodynamic limit of random regular graphs), and $\hG(z)$ the similar quantity in the absence of one incident edge. These
are random variables, because of the randomness in the choice of the
$H_{ii}=\epsilon_i$. The above written equations implies that their
distribution follows from
\begin{equation}
G(z) \eqd \frac{1}{\epsilon-z-\sum_{i=1}^{k+1} \hG_i(z)} \ , \ \ \
\hG(z) \eqd \frac{1}{\epsilon-z-\sum_{i=1}^k \hG_i(z)} \ ,
\label{eq_distrib}
\end{equation}
where $\eqd$ denotes the equality in distribution of two random variables.
In the r.h.s. the various $\hG_i$ are i.i.d. copies of $\hG$, and
$\epsilon$ is independently drawn uniformly on $[-W/2,W/2]$. The spectral
properties of the model can be obtained from the solution of these
distributional equations, by computing
\begin{equation}
\rho(E) = \lim_{\eta \to 0} \frac{1}{\pi} \Imp \E [ G(E+i\eta)] , \ \ \
L(E) = \lim_{\eta \to 0} \frac{1}{\pi} \eta \, \E [ |G(E+i\eta)|^2] \ ,
\label{eq_observables}
\end{equation}
where the expectation $\E[\cdot]$ is with respect the distribution of $G$.
The first quantity gives the density of states of energy $E$, while the second
one corresponds to the average inverse participation ratio of the eigenstates of
energy close to $E$ (for a finite graph, the inverse participation ratio of the eigenstate $\alpha$ is $L^\alpha = \sum_i |\psi_i^\alpha|^4$, where $\psi_i^\alpha$ is the component of the normalized $\alpha$-th eigenvector on site $i$): it
is non-zero only if $E$ falls in an interval of energy comprising
localized states (pure-point spectrum) and vanishes for extended states
(absolutely continuous spectrum). Indeed the density of states alone is
smooth at the mobility edge separating localized and extended
states~\cite{Wegner} and cannot be used to distinguish the two regimes. An
alternative and numerically more precise method to compute the mobility
edge was proposed in ref.~\citen{Abou1} and consists in investigating the
stability of a real solution of the equations (\ref{eq_distrib}) for real
$z=E$ to the insertion of a small imaginary part; this solution is stable
whenever $E$ belongs to the localized regime. Physically, for a finite and
large tree, this amounts to couple boundary sites to a thermal
bath and test whether the site at the center can dissipate energy at very
low temperature when the Fermi energy is close to $E$.
If $E$ belongs to the localized regime, energy is not transported across
the tree and no dissipation is possible far from
the boundary, i.e. the imaginary part of the self energy vanishes for
sites very distant from the boundary.

A numerical procedure to solve the distributional equations
(\ref{eq_distrib}) was proposed in ref.~\citen{Abou1}, and revivified more
recently in the context of finite-connectivity mean-field disordered
systems under the name of population dynamics~\cite{MePa_Bethe}, also
called the pool method~\cite{Monthus}. The idea is to approximate the
distribution of a random variable, say $\hG$, by the empirical
distribution of a sample of a large number $\N$ of representants
$\{\hG_1,\dots,\hG_\N\}$. Starting from an arbitrary initial condition a
sequence of samples is produced. The elements of the new sample $\hG'_j$
are generated by, identically and independently for each $j$, selecting
$k$ elements uniformly at random from the current sample
$\{\hG_1,\dots,\hG_\N\}$, drawing an energy $\epsilon$ uniformly in
$[-W/2,W/2]$, and computing the value of $\hG'_j$, according to
(\ref{eq_distrib}). Repeating these steps, one reaches a sample
approximating the fixed point solution of (\ref{eq_distrib}), and then the
observables in (\ref{eq_observables}) can be obtained by computing
empirical averages over the representants of the distribution. The
numerical accuracy is controlled by the size $\N$ of the samples.
\section{Phase diagram, density of states and mobility edges}
\label{sec_pd}
\begin{figure}
\centerline{\includegraphics[width=9cm]{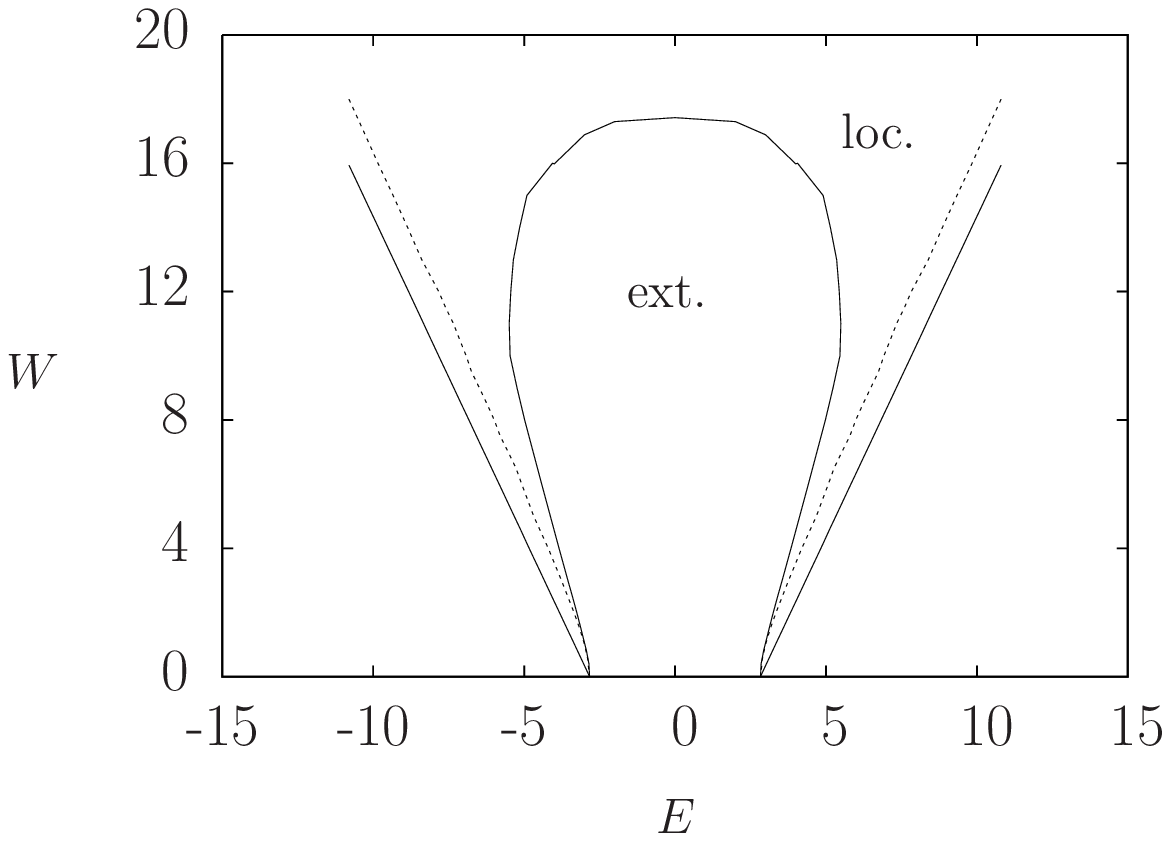}}
\caption{Phase diagram for the Bethe lattice with connectivity $k+1=3$.
The innermost solid line indicates the mobility edge between extended and
localized states, the outermost solid line being the edge of the density
of states $E=\pm(2\sqrt{k}+W/2)$. The dashed line is the numerically
observed edge, see the text for details.}
\label{fig_phasediagram}
\end{figure}

We have numerically solved the equations (\ref{eq_recurs}) with the method
described above for the connectivity $k+1=3$ and using samples with sizes
up to $\N=10^7$. The results are displayed in Fig.~\ref{fig_phasediagram}.
The innermost solid line is the mobility edge separating the regimes of
extended and localized states. It crosses the $E=0$ axis for a critical
value of the disorder $W_{\rm c} \approx 17.4$, above which no extended
states exist. This value is in agreement with the one found in
refs.~\citen{Abou1,Abou2,Monthus}. Moreover for small disorder the whole
interval $[-2\sqrt{k},2\sqrt{k}]$ of the density of states of the pure model ($W=0$) corresponds to extended states, in other words the mobility
edge does not enter the band of the pure model up to a strictly positive value of $W$. This fact has indeed been proven
rigorously~\cite{Klein1,ASW}. The dashed lines show the numerically
determined edge for the density of states, that is the limit of the
interval for which $\rho(E)$ of Eq.~(\ref{eq_observables}) is found to be
non-vanishing. The edge of the spectrum has been proved rigorously
to be equal to $E=\pm(2\sqrt{k}+W/2)$ (solid lines in
Fig.~\ref{fig_phasediagram}) by ergodicity
arguments~\cite{Klein1,Klein2,ASW}. The discrepancy between the
numerically computed value and the true one is due to the extreme
smallness of the density of states close to the edge. As we shall discuss
in the next section, the decrease of $\rho(E)$ is extremely fast.
Therefore, one would need extremely large sample sizes $\N$ in order to
reproduce the density of states close to the edge. We have checked that in
agreement with this interpretation the location of the observed band edge
depends on the sample size $\N$ used. The data presented in the figure for
the band edge corresponds to $\N=2 \, 10^6$.

\begin{figure}
\centerline{\includegraphics[width=9cm,angle=270]{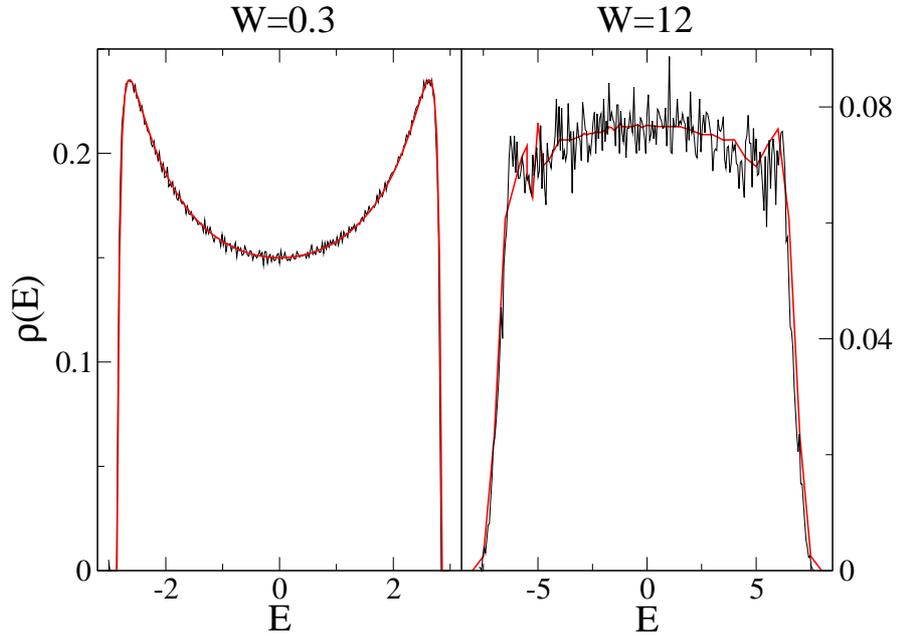}}
\caption{Density of states, $\rho(E)$, obtained using the cavity method
and exact diagonalization of the connectivity matrix of random regular graphs
of $8192$ sites for $W=0.3$ (left panel) and $W=12$ (right panel).
The data are averaged over $16$ different realizations
of the disorder and of the graph.}
\label{fig_spettro}
\end{figure}

\begin{figure}
\centerline{\includegraphics[width=9cm,angle=270]{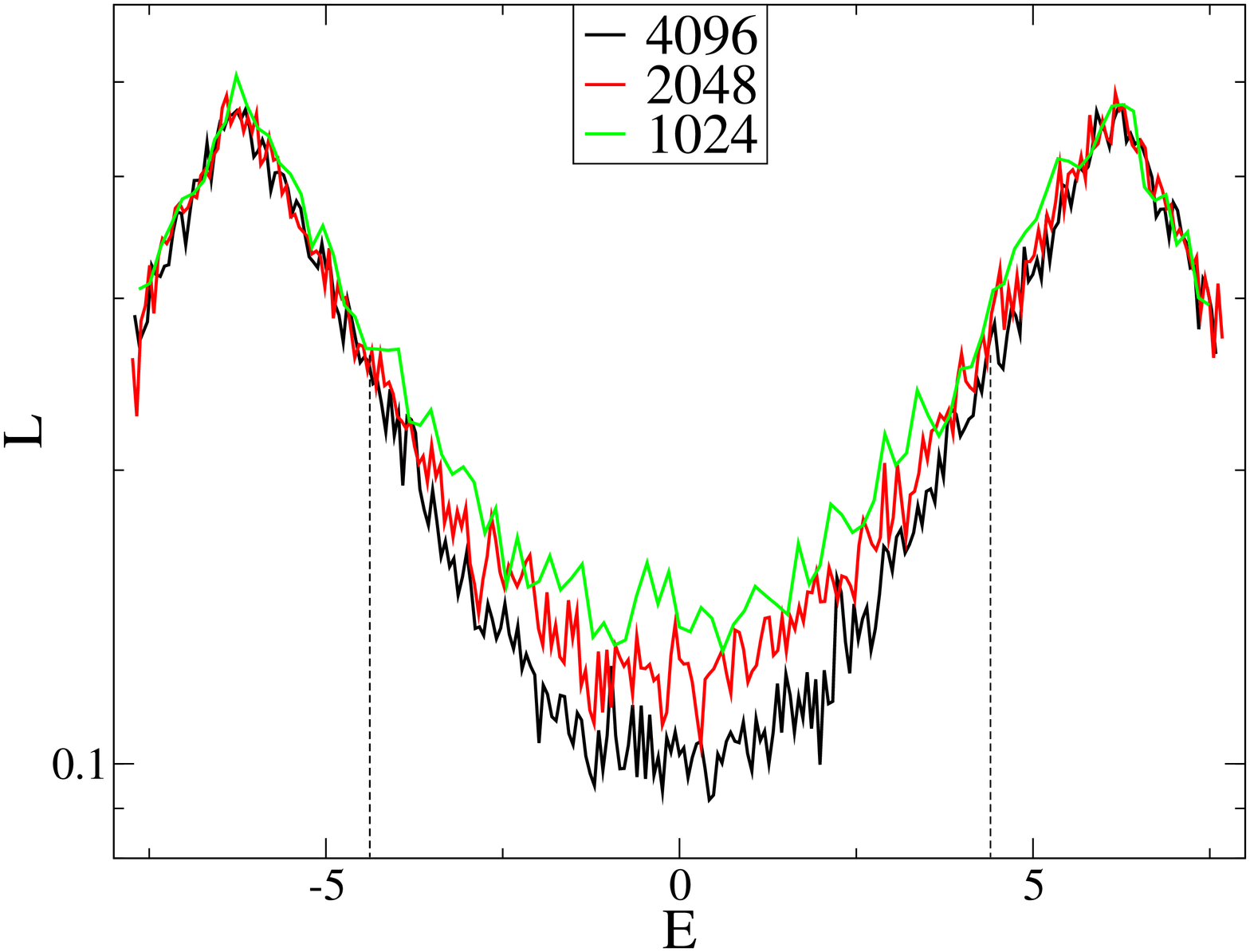}}
\caption{Inverse participation ratio, $L$, as a function of the energy
obtained via exact diagonalization for $W=12$ and for different sizes of the
adjacency matrix ($1024, 2048, 4096$). In the localized regime $L$ does not
depend on the size of the matrix, whereas the inverse participation ratio
of the extended states decrease as the inverse of the system size.
The mobility edge is found in $E \simeq 4$, in agreement, within the
numerical accuracy, with the numerical prediction of the cavity method.}
\label{IPR}
\end{figure}

The shape of the density of states and its evolution with the disorder is
shown in Fig.~\ref{fig_spettro}, where we plotted $\rho(E)$ for $W = 0.3$
and $W = 12$.
We also show the density of states obtained by exact diagonalization
for matrices of sizes $8192$. The agreement between the two methods is very good. However, for
strong disorder finite size effects become important close to the edge,
as expected from the discussion above. In Fig.~\ref{IPR} we show the inverse participation
ratio $L^\alpha$ obtained by exact diagonalization (we recall that for a finite
system $L^\alpha= \sum_i |\psi_i^\alpha|^4 $, where $\psi_i^\alpha$ is the value
of the normalized $\alpha$-th eigenvector at site $i$). The evolution of the
inverse participation ratio with the system size points to a mobility edge
$E_c\simeq 4$ for $W=12$, which is in agreement with the
prediction of the cavity method, see Fig.~\ref{fig_phasediagram}.

We have also studied the evolution of these results with the connectivity
of the Bethe lattice. The phase diagrams are qualitatively similar. The
main issue is the evolution of the mobility edge with $k$. In particular
we have focused on
the critical value of the disorder, $W_{\rm c}$, above which only
localized states exist. This is interesting for several reasons.
First, it is relevant for recent works studying the effect of
electron-electron interaction on Anderson localization.
In this case one is interested in the problem of localization in the Fock
space, which has been argued to be related to the localization problem
on a Bethe lattice with very large connectivity \cite{altshuler}. Second,
the limit of large connectivity might be the only case
where a complete analytical solution is within reach. Abou-Chacra et al.
\cite{Abou1} developed two different approximation schemes to obtain the
mobility edges analytically. Both methods lead to an asymptotic form of
$W_{\rm c}$ of the form $c\, k \ln k$. The constants $c$ are equal to $4$
and $2e\simeq 5.43$, depending on the approximation.
We have found that the stability analysis used to determine the mobility
edge can be interpreted, in the large $k$ limit, as an iterative equation
on Levy random variables with tail exponent $\mu=1/2$. This leads to the
result $c=2\sqrt{2\pi}\simeq 5.01$.
Whether this is still an approximation or an exact result is unclear at
this stage and will be investigated and
detailed elsewhere \cite{bst}.

Our numerical results for $W_{\rm c}$ obtained by solving the distributional equation (\ref{eq_distrib}) are
presented in table \ref{tab1}.
None of the analytical predictions cited above fit well the data.
Presumably, very strong sub-leading corrections
are present, possibly scaling like $k$ with potentially $\ln \ln $
corrections. In this situation one would need very large
values of $k$ to find the true asymptotic form. A best fit of our data of
the form $W_{\rm c} = c\, k \ln k +b\, k$ leads to $c=4.7$ and $b=5.9$.
Imposing one of the three values of $c$ quoted above and adjusting the parameter
$b$ leads to fits of comparable quality, these numerical data are hence
insufficient to discriminate between the proposed values of $c$.

\begin{table}
\begin{center}
\begin{tabular}{|c|c|c|c|c|c|c|c|}
\hline
$k+1$ & 3 & 4 & 5 & 6 & 7 & 8 & 9\\
\hline
$W_{\rm c}$ & 17.4  &  33.2 & 50.1 & 67.7 &  87.3 & 105 & 125.2\\
\hline
\end{tabular}
\end{center}
\caption{Critical value of the disorder as a function of the connectivity.}
\label{tab1}
\end{table}

\section{Lifshits tail and the edge of the spectrum. }

As discussed in the previous section, it was proved rigorously that the
edge of the density of state is located in $E = \pm
(2 \sqrt{k} + W/2)$~\cite{Klein1,Klein2,ASW}. Numerically we found that
the density of states becomes extremely small close to the edges.
In the following we shall develop simple arguments \`a la Lifshits that
allows one to understand
the result $E = \pm (2 \sqrt{k} + W/2)$ and to obtain the form of
$\rho(E)$ close to the edge.

Let us first recall how the Lifshits' argument~\cite{Lifshits} works on
finite dimensional lattices. Let us focus on
the left edge (the argument for the right edge is the same). The key
observation is that
there is a finite probability to have all the on-site random energies
inside a sphere of radius $R$ taking values very close to
$-W/2$, say $-W/2+\epsilon$. Using that such a sphere, when disconnected
from the rest, is characterized
by states with energies arbitrarily close to the exact band edge in absence
of disorder shifted by $-W/2+\epsilon$,
one can construct variational states for the original Hamiltonian that
have the same energies.
This allows one to show that the left and right edges in presence of
disorder are just the ones
without disorder shifted respectively by $-W/2$ and $W/2$, and also to
find the decrease of $\rho(E)$ at the edges. Note that all this can be put
on a rigorous ground, as done for instance in ref.~\citen{book_Pastur}.

In this section we shall develop similar arguments to study the behaviour
of the tail of
the spectrum of the Anderson model on Bethe lattices (a related computation
was presented in ref.~\citen{KH} for off-diagonal disorder).
Following the strategy developed for finite dimensional systems,
the first thing to analyze is the eigenvalue problem for an infinite tree
of connectivity $k+1$ without disorder, i.e.
where the adjacency matrix $H$ is equal to $-1$ if $i$ and $j$ are nearest
neighbours and zero otherwise.
We look for eigenvectors of $H$ of energy $E$, i.e., we
want to find
states $|\psi \rangle$  such that $H |\psi \rangle = E |\psi \rangle$.
We consider spherically symmetric states for which the components of $|\psi \rangle$ depend
only on the distance from the central site of
the tree and must satisfy the following set of equations:
\begin{eqnarray} \label{var1}
E \psi_n & = & - k \psi_{n+1} - \psi_{n-1} \\
\nonumber
E \psi_0 & = & - (k+1) \psi_1
\end{eqnarray}
where $\psi_n$ is the component of $|\psi \rangle$ on any given site at
distance $n$ from the origin.
For energies in the range $-2 \sqrt{k} < E < 2 \sqrt{k}$,
the solution of the equations above can be written in the form:
\begin{equation}
\psi_n = k^{-n/2} \left( \frac{1-ic}{2} \, e^{i n \theta} + \frac{1+ic}{2}
\, e^{-i n \theta} \right) \ ,
\end{equation}
with $e^{i \theta} = (-E + i \sqrt{4 k - E^2})/2 \sqrt{k}$.
The value of $c$ must be chosen in order to satisfy the equation for the
central site. This yields:
\begin{equation}
c = - %\frac{\sqrt{k}}{\sin \theta}\left(\frac{E}{k+1} +\frac{\cos \theta}{\sqrt k}\right)
\frac{(k-1) E}{(k+1) \sqrt{4k - E^2}} \ .
\end{equation}
Finally, the component of the eigenvector $| \psi \rangle$ at distance $n$
from
the origin can be rewritten as:
\begin{equation}
\psi_n = \frac{\cos(n \theta) + c \sin (n \theta)}{k^{n/2}} \ .
\end{equation}
Thus $| \psi_n |^2$ decrease with the distance from the central site
as $k^{-n}$, while the number of sites at distance $n$ from the origin grows
as $k^n$. As a consequence, such radial (quasi-)eigenvectors are not
normalizable:
\begin{equation}
\langle \psi | \psi \rangle = 1 + \frac{k+1}{k} \sum_{n \ge 1} k^n |
\psi_n |^2
= 1 + \frac{k+1}{k} \sum_{n\ge 1} | \cos(n \theta) + c \sin (n \theta) |^2
\ ,
\end{equation}
which diverges for an infinite tree.

One can however define a modified variational wave-function that is normalizable
and has variational energy equal to $E$. The idea is the following: define
$n_\star(E)$ as the smallest integer $n$
such
that $\psi_n=0$, and consider the truncated wave-function $\psi'_n=\psi_n$ if
$n\le n_\star(E)$, $\psi'_n = 0$ for $n\ge n_\star(E)$. This wave-function
is not a
solution of the equations (\ref{var1}), yet it is normalizable and verifies
$\frac{\la \psi'|H|\psi'\ra}{\la \psi'|\psi' \ra}=E$. The determination of
the radius $n_\star(E)$ is simplified when $E$ tends to one of the band-edge.
Let us focus on the left band edge (the argument for the right edge is almost identical) and write $E=-2\sqrt{k}+\delta$. At the lowest order in $\delta$ one finds
that $\theta \simeq \sqrt{\frac{\delta}{\sqrt k}}$ and
$c \simeq \frac{k-1}{k+1} \sqrt{\frac{\sqrt k}{\delta}}$. Solving the equation
$\psi_{n_\star}=0$ in this limit one obtains
$n_\star \sim \pi \sqrt{\sqrt{k}/\delta}$.

Let us now go back to the disordered case, where the on-site energies
$H_{ii}$ are independent and identically distributed random variables
in the interval $[-W/2,W/2]$, and let us use the variational states
derived above to determine the position of the edge of the
spectrum in presence of disorder.
There is a finite probability that inside a spherical region of radius
$n_{\star}$ all the on-site energies $\epsilon_i$ are arbitrarily close to $-W/2$. The variational state constructed above has now a variational energy arbitrarily
close to $-2\sqrt{k}-W/2$. The same argument for the right edge leads to the $2\sqrt{k}+W/2$. As a consequence, the edges of the Anderson model on the Bethe lattice are $\pm(2\sqrt{k}+W/2)$, in agreement with the rigorous results based on
ergodicity theorem~\cite{Klein1,Klein2}.

One can make this argument more quantitative and estimate the
behaviour of the density of states around the left edge of the band, in
presence of disorder (again the argument for the right edge is essentially identical). Indeed, the probability that all the energies in
the ball of radius $n_\star$ are in the interval $[-W/2,-W/2+\delta]$
scales as $(\delta/W)^{\left(\frac{k+1}{k} \right)k^{n_\star}}$, since the number of sites inside the ball is
$\left(\frac{k+1}{k} \right)k^{n_\star}$. Combining this relation with the scaling of $n_\star$
with $\delta$ obtained above for the ordered model yields the following
scaling for the density of states
\begin{equation}
\rho(-2\sqrt{k}-W/2+\delta) \simeq \exp\left[-\left(\frac{k+1}{k} \right) \, k^{\pi k^{1/4} \, \delta^{-1/2}}
\log(W/\delta) \right] \ \label{eq:lifschitz}
\end{equation}
This function vanishes indeed extremely
fast with $\delta$, as mentioned previously when discussing the numerical
results of Sec.\ref{sec_pd}. Note that, strictly speaking, this expression is only a lower bound to the density of states. A form similar to Eq.~(\ref{eq:lifschitz}) has already been
found for models with off-diagonal disorder on Bethe
lattices~\cite{KH,tail_Bethe}.

\section{Isolated eigenvalue}
\label{sec_isolato}
The last point we shall address is whether the smallest eigenvalue
$\lambda_1$ is at a finite distance from the edge of the
density of states at the thermodynamic limit. Note that one needs an
extensive number of eigenvalues to obtain a $\rho(E)$ different from zero.
A single eigenvalue gives a contribution to $\rho(E)$ of the order of
$1/N$, which is negligible in the thermodynamic limit.
As a consequence, it is in principle possible to have the smallest
eigenvalue separated from all the others by a gap and the left edge of $\rho(E)$ at a finite distance
from $\lambda_1$.
Although this might seem counterintuitive at first sight, at least based
on the intuition
developed for finite dimensional system, it is what happens in the case
without disorder when $H$ is (minus) the adjacency matrix of a random
regular graph\footnote{For this phenomenon, there is a difference between
an infinite random tree and a very large random regular graph. Only for the
latter
the smallest eigenvalue is separated from the others by a gap.}. It is easy
to check in this case that the constant vector is an eigenvector of $H$
with eigenvalue $-(k+1)$. It appears thus a gap between $\lambda_1=-(k+1)$
and the second smallest eigenvalue $\lambda_2$, which concentrates around
$-2\sqrt{k}$, the left edge of the band of the Bethe lattice model, as
proved in ref.~\citen{Friedman}.

A natural question is therefore what happens to the isolated smallest
eigenvalue when disorder is introduced.
It is also interesting to investigate the properties of the corresponding
eigenvector, which it is completely delocalized
without disorder. We shall follow an approach first proposed in
ref.~\citen{giap} for off-diagonal disorder, and begin by rederiving it
with a slightly different presentation, related to the discussion of
ref.~\citen{almost}. Let us consider a Gaussian probability measure on $N$
real variables $\phi_i$ defined by the averages
\begin{equation}
\la \bullet \ra = \frac{1}{\cal N}\int \dd \phi_1 \dots \dd \phi_N \bullet
\exp\left[-\frac{1}{2}\sum_{i,j} \phi_i ( H_{i,j} - E \delta_{i,j} )
\phi_j\right] \ ,
\end{equation}
where $\cal N$ is a normalizing factor. It is clear that the integral is
convergent only if $E<\lambda_1$, and that by symmetry $\la \phi_i \ra=0$
for all $i$. If however the limit $E\to \lambda_1^-$ is taken then an
infinitesimal linear term in the action above is enough to produce a
spontaneous magnetization $\la \phi_i \ra \neq 0$ in the direction of the
eigenvector associated to $\lambda_1$, i.e. $\la \phi_i \ra$ is
proportional to its $i$-th component.
This corresponds to a condensation transition on the first eigenvector.
Using the replica symmetric cavity method~\cite{MePa_Bethe} (which
corresponds here to the Gaussian Belief Propagation~\cite{GBP}) one finds
that computing the averages with respect to this weight, on a given graph,
amounts to solve the following recursion equations
\begin{equation}
\mu_{i \to j}(\phi_i) = \frac{1}{{\cal N}_{i \to j}}
e^{-\frac{1}{2}(H_{ii}-E) \phi_i^2}
\prod_{j' \in \partial i \setminus j} \int \dd \phi_{j'} \mu_{j' \to
i}(\phi_{j'}) e^{-H_{ij'} \phi_i \phi_{j'}} \ ,\label{eq_recurs2}
\end{equation}
where ${\cal N}_{i \to j}$ ensures the normalization and $\mu_{i \to
j}(\phi_i) $ is the probability measure for $\phi_i$
when the link between $i$ and $j$ is removed\footnote{The replica symmetric cavity method consists in assuming
that the $\phi_{j'}$ are not correlated except through their coupling to $i$. This is justified in the present case and directly leads to the equations (\ref{eq_recurs2}).}. We parametrize this Gaussian
probability as:
\begin{equation}
\mu_{i \to j}(\phi_i) = \frac{1}{{\cal N}'_{i \to j}}
e^{-\frac{1}{2}\frac{1}{G_{i \to j}} \phi_i^2 - y_{i \to j} \phi_i} \ .
\end{equation}
Plugging this expression in (\ref{eq_recurs2}) we find:
\begin{equation}
G_{i \to j} = \frac{1}{H_{ii}-E-\underset{j' \in \partial i \setminus
j}{\sum} H_{ij'}^2 G_{j' \to i}} \ ,
\ \ \ y_{i \to j} = - \sum_{j' \in \partial i \setminus j} H_{ij'} G_{j'
\to i} y_{j' \to i} \ ,
\end{equation}
which are indeed the same as derived in ref.~\citen{giap}. The equations
linking the 'messages' $G$ on the various edges of the graph are decoupled
from the $y$'s, and corresponds to the recursion equations for the
resolvents stated in (\ref{eq_recurs}). The linear term coefficients $y$
are solutions of a homogeneous set of equations, hence invariant with
respect to a common multiplication. The smallest eigenvalue $\lambda_1$
can thus be found as the largest value of $E$ such that the fixed point
$y=0$ is stable.

To investigate the behaviour of $\lambda_1$ for large regular graphs one
turns these equations in distributional equations,
\begin{equation}
(\hG,y) \eqd \left( \frac{1}{\epsilon - E - \sum_{i=1}^k \hG_i} ,
\sum_{i=1}^k \hG_i y_i \right) \ .
\label{eq_iso_distrib}
\end{equation}
As explained in Sec.~\ref{sec_model} this kind of distributional equation
is easily solved numerically by representing the distribution of the
random variable $(\hG,y)$ as a sample of couples
$\{(\hG_1,y_1),\dots,(\hG_\N,y_\N)\}$. These are updated according to a
generalization of the method explained in Sec.~\ref{sec_model}, where now
the update is made on the couple $(\hG_j,y_j)$ and not on $\hG_j$ alone.
The stability of the $y=0$ fixed point can then be determined by
monitoring the norm $\sum_j y_j^2$ after each iteration, and one
identifies the location of $\lambda_1$ as the largest value for which this
norm decreases upon iterating Eq.~(\ref{eq_iso_distrib}).

\begin{figure}
\centerline{\includegraphics[width=9cm]{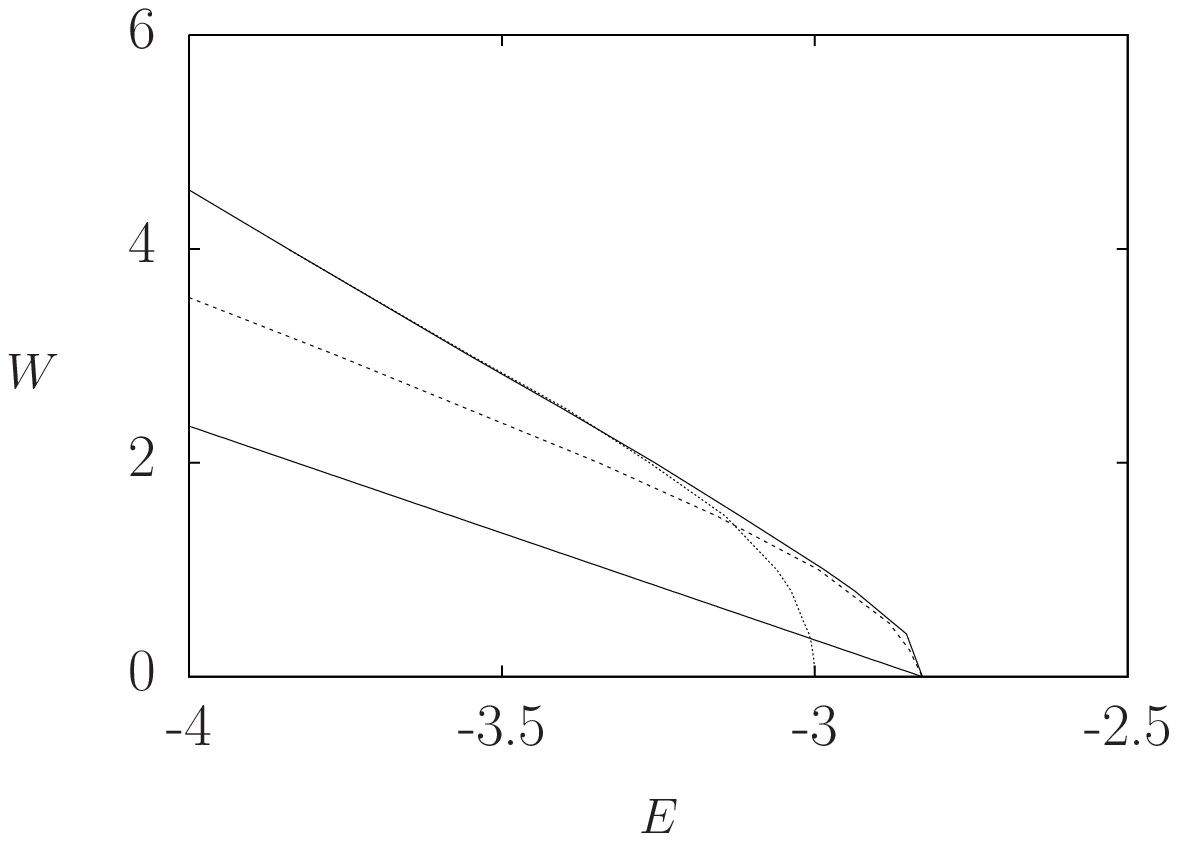}}
\caption{Blow-up of the phase diagram of Fig.~\ref{fig_phasediagram}, the
supplementary line starting from $-3$ is the isolated eigenvalue
$\lambda_1$.}
\label{fig_isolato}
\end{figure}

The numerical results of this procedure are presented in
Fig.~\ref{fig_isolato} for the connectivity $k+1=3$. The smallest
eigenvalue $\lambda_1$ sligthly decreases upon increasing the disorder
from the value $-3$ it has at $W=0$.
We find that even in presence of disorder the smallest eigenvalue is
separated from the others by a gap.
Moreover it remains delocalized as shown by its participation ratio that
can be obtained from the $y_{i \to j}$ when $E=\lambda_1^-$.
There is a critical value of the disorder, $W_1$, where $\lambda_1$
becomes equal to the edge of the density of states.
For larger values of $W$ the smallest eigenvalue equals the edge of the
density of states\footnote{
For $W>W_1$ our method still finds a value of
$\lambda_1$, which is now
{\it larger} then the left band edge and that approaches the mobility edge
when increasing the disorder strength.
The interpretation of the replica symmetry cavity method in this regime is
unclear. It might be that
between the extensive number of localized states, there is one extended
state at least for $W$ not much larger than $W_1$
and that the  replica symmetry cavity method misses all the other
localized states because looking at the stability
with respect to $y_{i \to j}$ is like studying the stability to a
delocalized external field, which does not couple to localized states.
This point certainly deserves further investigation.}. In Fig.
\ref{fig_isolato} we show the numerical result obtained by the cavity
method
for the isolated eigenvalue in the case $k+1=3$, which crosses the left
edge of the band for a value of the disorder $W_1\simeq 0.36$.

\section{Conclusion}
In this work we have revisited the Anderson localization problem on a
Bethe lattice.
We have shown detailed results on the density of states and the mobility
edge for connectivity equal to $3$.
We have also discussed the large connectivity limit, whose exact solution
remains an open question.
Finally, we have shown that for not too large disorder the smallest
eigenvalue is extended and has a gap
from the subsequent (localised) eigenvalues. This property may play an
important role for the disordered Bose
Hubbard model on Bethe lattice. Indeed, the usual argument used to claim
that a Bose glass phase
always intervenes between the superfluid and the Mott phase may fail. In
finite dimensional lattices,
when the disorder strength $W$ is only slightly larger than the gap of the Mott
insulator, it was argued~\cite{dbh,proko} that it is favorable to
insert particles (or vacancies) but that these are not delocalized and
hence do not lead to superfluidity since
the lowest eigenvalue of the corresponding Anderson problem is localized.
In the Bethe lattice case, depending
on the effective Anderson model one finds for these quasi-particles, the
smallest eigenvalue may correspond
to a completely delocalized state with a finite gap from the localized
ones. Thus, the possibility of a direct quantum phase
transition from the superfluid to the Mott state is not completely
excluded on Bethe lattices.

\section*{Acknowledgements}

We would like to thank M. Aizenman, J.P.~Bouchaud, M. Lelarge and F. Zamponi for useful discussions.

%\appendix
%\section{First Appendix} %Empty argument \section{} yields `Appendix'.
%
%\section{Second Appendix}

\end{document}